\documentclass[runningheads]{llncs}
\usepackage[utf8]{inputenc}
\usepackage{amsmath}
\usepackage{amsfonts}
\usepackage{amssymb}
\usepackage{graphicx}
\usepackage{algorithm2e}
\usepackage{cite}
\usepackage{siunitx}
\usepackage{chngcntr}
\usepackage{tikz}
\usepackage{tkz-graph}
\usepackage{epsfig}
\usepackage{epstopdf}
\usepackage{hyperref}
\usepackage{bookmark}
\usepackage{xcolor,colortbl}
\usepackage[obeyFinal]{todonotes}

\usepackage{url}
\urldef{\mailsa}\path|{chromy}{kucerap}@ktiml.mff.cuni.cz|
\urldef{\mailsb}\path|kucerap@ktiml.mff.cuni.cz|

\makeatletter
\renewcommand*{\thetable}{\arabic{table}}
\renewcommand*{\thefigure}{\arabic{figure}}
\let\c@table\c@figure%
\makeatother
\begin{document}
\title{Phase Transition in Matched Formulas and a~Heuristic for
   Biclique
   Satisfiability\thanks{This research was supported by SVV project number 260 453.
   Access to computing and storage facilities owned by parties and projects contributing to the National Grid Infrastructure MetaCentrum provided under the programme "Projects of Large Research, Development, and Innovations Infrastructures" (CESNET LM2015042), is greatly appreciated.} }

 \author{{Miloš Chromý}\inst{1}    \and {Petr Kučera}\inst{1}}
\institute{Charles University, Faculty of Mathematics and Physics, Department of Theoretical Computer Science and Mathematical Logic, Malostransk{\'e} n{\'a}m. 25, 118\,00 Praha 1, Czech Republic, \email{\mailsa}}

\authorrunning{M. Chromý et al.}
\titlerunning{Biclique satisfiability}

   \maketitle

   \begin{abstract}
     A matched formula is a CNF formula whose incidence graph admits a matching which matches a distinct variable to every clause. We study phase transition in a context of matched formulas and their generalization of biclique satisfiable formulas. We have performed experiments to find a phase transition of property ``being matched'' with respect to the ratio \(m/n\) where \(m\) is the number of clauses and \(n\) is the number of variables of the input formula \(\varphi\). We compare the results of experiments to a theoretical lower bound which was shown by Franco and Gelder (2003). Any matched formula is satisfiable, moreover, it remains satisfiable even if we change polarities of any literal occurrences. Szeider (2005) generalized matched formulas into two classes having the same property --- var-satisfiable and biclique satisfiable formulas. A formula is biclique satisfiable if its incidence graph admits covering by pairwise disjoint bounded bicliques. Recognizing if a formula is biclique satisfiable is NP-complete. In this paper we describe a heuristic algorithm for recognizing whether a formula is biclique satisfiable and we evaluate it by experiments on random formulas. We also describe an encoding of the problem of checking whether a formula is biclique satisfiable into SAT and we use it to evaluate the performance of our heuristic.
       \keywords{SAT, matched formulas, biclique SAT, var-SAT, phase transition, biclique cover.}   
       
       \end{abstract}

   \section{Introduction}

   In this paper we are interested in the  \emph{problem of
      satisfiability} (SAT) which is central to many areas of
   theoretical computer science. In this problem we are given a formula
   \(\varphi\) in propositional logic and we ask if this formula is
   satisfiable, i.e.\ if there is an assignment of values to variables which satisfies \(\varphi\).
   This is one of the best known NP-complete
   problems~\cite{Cook:1971:CTP:800157.805047}. In this paper we
   study special classes of formulas whose definition is based on the notion of incidence graph.

   Given a
   formula \(\varphi\) in \emph{conjunctive normal form} (CNF) we
   consider its {\em incidence graph\/} $I(\varphi)$ defined as
   follows. $I(\varphi)$ is a bipartite graph with one part consisting
   of the variables of \(\varphi\) and the other part consisting of the
   clauses
   of $\varphi$. An edge $\{x,C\}$ for a variable $x$ and a clause $C$
   is in $I(\varphi)$ if $x$ or $\overline{x}$ appear in $C$.
   It was
   observed by the authors of~\cite{AL86} and~\cite{T84} that if $I(\varphi)$
   admits a matching of size
   $m$ (where $m$ is the number of clauses in $\varphi$), then
   $\varphi$ is satisfiable. Later the formulas satisfying this
   condition were called {\em matched formulas\/}
   in~\cite{Franco2003177}. Since a matching of maximum size in a
   bipartite graph can be found in polynomial time (see e.g.~\cite{HK73,LP86}), one can check efficiently whether a given formula is
   matched.


   It is clear that if \(\varphi\) is a formula on \(n\) variables and
   \(m\) clauses then \(\varphi\) can be matched only if \(m\leq n\).
   The authors of~\cite{Franco2003177} asked an interesting question:
   What is the probability that a formula \(\varphi\) is matched
   depending on the ratio \(\frac{m}{n}\)? We can moreover ask if
   the property ``being matched'' exhibits a phase transition.

   A phase transition was studied in context of
   satisfiability~\cite{Cheeseman:1991:RHP:1631171.1631221,Gent:Walsh:1994,Mitchell_et_al:1992,Dubois:2000:TRF:338219.338243,DBLP:journals/corr/abs-1202-0042}.
   The so-called satisfiability threshold for a given \(k\) is a value
   \(r_k\) satisfying the following property: A random formula
   \(\varphi\) in \(k\)-CNF on \(n\) variables and \(m\) clauses is
   almost surely satisfiable if \(\frac{m}{n}< r_k\) and it is almost
   surely unsatisfiable if \(\frac{m}{n}> r_k\). For instance the value
   \(r_3\) is approximately
   \(4.3\)~\cite{Dubois:2000:TRF:338219.338243,DBLP:journals/corr/abs-1202-0042}.

   In the same sense we can study threshold for property ``being
   matched''. It was shown in~\cite{Franco2003177} that a \(3\)-CNF
   \(\varphi\) on \(n\) variables and \(m\) clauses is almost surely
   matched if \(\frac{m}{n}< 0.64\).
   This is merely a theoretical lower bound, and in this paper we
   perform experimental check of this value. It
   turns out that the experimentally observed threshold is much higher
   than the theoretical lower bound. Moreover we observe that the property
   ``being matched'' has a sharp threshold or phase transition as a
   function of ratio \(\frac{m}{n}\).

   Matched formulas have an interesting property: If a formula
   \(\varphi\) is matched then we pick any
   occurrence of any literal and switch its polarity (i.e.\ change a
   positive literal \(x\) into a negative literal \(\overline{x}\) or
   vice versa). The formula produced by this operation will be matched
   and thus satisfiable as well. This is because the definition of incidence graph completely ignores
   the polarities of variables.
   The formulas with this property were called \emph{var-satisfiable}
   in~\cite{Szeider2005} and they form a much bigger class than matched formulas.
   Unfortunately, it was shown in~\cite{Szeider2005} that the problem
   of checking whether a
   given formula \(\varphi\) is var-satisfiable is complete for the
   second level of polynomial hierarchy.

   Szeider in~\cite{Szeider2005} defined a subclass of var-satisfiable
   formulas called \emph{biclique satisfiable formulas} which
   extends matched formulas. It was shown in~\cite{Szeider2005} that checking if
   \(\varphi\) is biclique satisfiable is an NP-complete problem. In this paper we
   describe a heuristic algorithm to test whether a formula is
   biclique satisfiable. Our heuristic algorithm is based on an
   heuristic for covering a bipartite graph with bicliques described
   in~\cite{conf/hicss/HeydariMSS07}. We test our heuristic algorithm
   experimentally on random formulas. Our heuristic algorithm is
   incomplete, in particular, whenever it finds that a formula is
   biclique satisfiable, then it is so, but it may happen that a
   formula is biclique satisfiable even though our algorithm is unable
   to detect it. In order to check the quality of our heuristic, we
   propose a SAT based approach to checking biclique satisfiability of a
   formula. We compare both approaches on random
   formulas.

   In Section~\ref{sec:defn} we recall some basic definitions and
   related results used in the rest of the paper. In
   Section~\ref{sec:matched} we give the results of experiments on matched
   formulas. In Section~\ref{sec:biclique} we describe our heuristic
   algorithm for determining if a formula is biclique satisfiable and
   we give the results of its experimental
   evaluation. In Section~\ref{sec:sat-encoding} we describe
   a SAT based approach to checking biclique satisfiability and compare it experimentally with the heuristic
   approach. We close the paper with 
   concluding remarks in Section~\ref{sec:concl} and give directions
   of further research in Section~\ref{sec:future}.

   \section{Definitions and Related Results}
      \label{sec:defn} 

     In this section we shall introduce necessary notions and results used in the paper.
     
      \subsection{Graph Theory}

      We use the standard graph terminology (see e.g.~\cite{B98}). A \emph{bipartite graph} \(G=(V_v, V_c, E)\) is a triple with vertices split into two parts \(V_v\) and \(V_c\) and the set of edges \(E\) satisfying that
      \(E\subseteq V_v\times V_c\). Given a bipartite graph \(G\) we
      shall also use the notation \(V_v(G)\) and \(V_c(G)\) to denote
      the vertices in the first and in the second part respectively.
      For two natural numbers \(n, m\) we denote by \(K_{n,m}\) the
      \emph{complete bipartite graph} (or a \emph{biclique}) that
      is the graph \(K_{n,m}=(V_v, V_c, E)\) with \(|V_v|=n\),
      \(|V_c|=m\) and \(E=V_v\times V_c\).

      Given a bipartite graph \(G=(V_v, V_c, E)\) the \emph{degree} of
      a vertex \(v\in V_v\cup V_c\) is the number of incident edges.
      A subset of edges $M\subseteq E$ is
      called a \emph{matching} of \(G\) if 
      every vertex in $G$ is incident to at
      most one edge in $M$. A vertex $v$ is \emph{matched} by
      matching $M$ if $v$ is incident to some edge from $M$.
      $M$ is a \emph{maximum matching} if for every other matching
      $M'$ of $G$ we have that $|M|\geq|M'|$. There is a polynomial
      algorithm for finding a maximum matching of a bipartite graph
      \(G=(V_v, V_c, E)\) which runs in
      $O(|E|\sqrt{|V_v|+|V_c|})$~\cite{HK73,LP86}.

      \subsection{Boolean Formulas}

      A \emph{literal} is a variable $x$ or its negation
      $\overline{x}$. A \emph{clause} is a finite disjunction of
      distinct literals $C=(l_1\vee l_2\vee \ldots \vee l_k)$, where $k$ is
      the \emph{width of clause} \(C\). A formula in
      \emph{conjunctive normal form} (\emph{CNF}) is a finite
      conjunction of clauses
      $\varphi=C_1\wedge C_2 \wedge\ldots\wedge C_n$. Formula \(\varphi\) is in
      $k$-CNF if all clauses in \(\varphi\) have width at most \(k\).
      We shall also often write (\(k\)-)CNF \(\varphi\) instead of
      \(\varphi\) in (\(k\)-)CNF\@. 


      Let us now recall the definition of probability
      space \(\mathcal{M}^k_{m,n}\) from~\cite{Franco2003177}.

      \begin{definition}[Franco and Van Gelder~\cite{Franco2003177}]
         Let \(V_n=\{v_1, \dots, v_n\}\) be a set of Boolean variables
         and let \(L_n=\{v_1, \overline{v_1}, \ldots, v_n,
               \overline{v_n}\}\) be the set of literals over
         variables in \(V_n\). Let \(\mathcal{C}^k_{n}\) be the set
         of all clauses with exactly \(k\) variable-distinct literals
         from \(L_n\). A random formula in probability space
         \(\mathcal{M}^k_{m,n}\) is a sequence of \(m\) clauses from
         \(\mathcal{C}^k_{n}\) selected uniformly, independently,
         and with replacement.
      \end{definition}

      \subsection{Matched Formulas}

      Let $\varphi=C_1\wedge\dots\wedge C_m$ be a CNF formula on $n$
      variables $X=\{x_1, \dots, x_n\}$. We associate a bipartite
      graph $I(\varphi)=(X,\varphi,E)$ with $\varphi$ (also called the
      \emph{incidence graph} of $\varphi$), where the vertices correspond to the
      variables in $X$ and to the clauses in $\varphi$. A variable \(x_i\)
      is connected to a clause $C_j$ (i.e.~$\{x_i,C_j\}\in E$) if
      $C_j$ contains $x_i$ or $\overline{x_i}$. A CNF formula $\varphi$
      is {\em matched\/} if $I(\varphi)$ has a matching of size
      $|\varphi|$, i.e.~if there is a matching which pairs each clause
      with a unique variable. It was observed in~\cite{AL86,T84} that a matched CNF is always satisfiable since
      each clause can be satisfied by the variable matched to the
      given clause. A variable which is matched to some clause in a
      matching $M$ is called {\em matched\/} in $M$.

      We can see that checking if a formula is matched amounts to
      checking if the size of a maximum matching of \(I(\varphi)\) is
      \(|\varphi|\). This can be done in time \(O(\ell\sqrt{m+n})\)
      where \(m\) denotes the number of clauses in \(\varphi\), \(n\)
      denotes the number of variables in \(\varphi\), and \(\ell\)
      denotes the total length of formula \(\varphi\) that is the sum
      of the widths of the clauses in \(\varphi\).

      The following result on density of matched formulas in the
      probability space \(\mathcal{M}^k_{m,n}\) was shown
      in~\cite{Franco2003177}.

      \begin{theorem}[Franco and Van Gelder~\cite{Franco2003177}]
         \label{thm:fg-matched} 
         Under \(\mathcal{M}^k_{m,n}\) \(k\geq 3\),
         the probability that a random formula \(\varphi\) is matched
         tends to \(1\) if \(r\equiv \frac{m}{n}<0.64\) as
         \(n\to\infty\).
      \end{theorem}

      One of the goals of this paper is to check experimentally how
      good estimate of the real threshold is the theoretical value \(0.64\).

      \subsection{Biclique Satisfiable Formulas}

      One of the biggest limitations of matched formulas is that if
      \(\varphi\) is a matched formula on \(n\) variables and \(m\)
      clauses, then \(m\leq n\). To overcome this limitation while
      keeping many nice properties of matched formulas, Stefan Szeider
      introduced biclique satisfiable formulas
      in~\cite{Szeider2005}.

      We say that a biclique \(K_{n,m}\) is bounded if \(m<2^n\).
      Let \(\varphi\) be a CNF on \(n\) variables and \(m\) clauses
      and let us assume that \(I(\varphi)=K_{n,m}\) where $m<2^n$.
      Then
      \(\varphi\) is satisfiable~\cite{Szeider2005}. This is
      because we have \(m<2^n\) clauses each of which contains
      all \(n\) variables. Each of these clauses determines
      one unsatisfying assignment of \(\varphi\), but there is \(2^n\)
      assignments in total. Thus one of these must be satisfying.

      Based on this observation we can define biclique satisfiable
      formulas~\cite{Szeider2005}. We say, that a bipartite graph $G=(V_v,V_c,E)$ has a
      \emph{bounded biclique cover} if there exists a
      set of bounded bicliques $\mathcal{B}=\{B_1,\ldots,B_k\}$
      satisfying the following conditions.
      \begin{itemize}
         \item every \(B_i, i=1,\ldots,k\) is a subgraph
         of G,
         \item for any pair of indices \(1\leq i<j\leq k\)
            we have that \(V_v(B_i)\cap V_v(B_j)=\emptyset\), and
         \item for every \(v\in V_c(G)\) there is a biclique \(B_i,
               i=1, \dots, k\) such that \(v\in V_c(B_i)\).
      \end{itemize}
      If every biclique $B_i\in\mathcal{B}$ in the cover satisfies that
      $|V_v(B_i)|\leq k$, then we say the graph $G$ has a bounded $k$-biclique cover.
      A formula $\varphi$ is \emph{($k$)-biclique satisfiable} if its
      incidence graph $I(\varphi)$ has a bounded ($k$-)biclique cover. 

      It can be easily shown that any biclique satisfiable formula is
      indeed satisfiable, however, it is an NP-complete problem to
      decide if a formula is biclique satisfiable even if we only
      restrict to \(2\)-biclique satisfiable formulas. For proofs of
      both results see~\cite{Szeider2005}. On the other hand it is
      immediate that $1$-satisfiable formulas are matched formulas,
      because a single edge is a bounded biclique.

      \subsection{Generating experimental data}
	\label{sec:generator}

      Whether a formula \(\varphi\) in CNF is matched or not depends
      only on its incidence graph \(I(\varphi)\). Instead of random
      formulas from probabilistic space \(\mathcal{M}^k_{m,n}\) we
      thus consider random bipartite graphs \(G=(V_v, V_c, E)\) from
      probabilistic space \(\mathcal{G}^k_{m,n}\).

      \begin{definition}
              Probability space \(\mathcal{G}^k_{m,n}\) is defined as
              follows. A random bipartite graph \(G\in\mathcal{G}^k_{m,n}\)
              is a bipartite graph with parts \(V_v, V_c\) where
              \(|V_v|=n\), \(|V_c|=m\). Each vertex \(v\in V_c\) has \(k\)
              randomly uniformly selected neighbours from \(V_v\).
      \end{definition}

      In our experiments we generated bipartite graphs $G\in \mathcal{G}_{m,n}^k$.
      Since we consider choosing clauses in formula \(\varphi\in \mathcal{M}^k_{m,n}\) with replacement, we can have several copies of the same clause in \(\varphi\). It follows that given a bipartite graph \(G\in\mathcal{G}^k_{m,n}\), we have exactly \(2^{k}m\) formulas \(\varphi\in\mathcal{M}^k_{m,n}\) which have \(I(\varphi)=G\) --- each vertex \(c\in V_c\) can be replaced with \(2^k\) different clauses with setting polarities to variables \(x\in V_v\) adjacent to \(v\) in \(G\). 
      In particular, the probability that a random formula
       \(\varphi\in\mathcal{M}^k_{m,n}\) is matched is the same as the
       probability that a random bipartite graph
       \(G\in\mathcal{G}^k_{m,n}\) admits a matching of size \(m\).
       The same holds for the biclique satisfiability. 



   \section{Phase Transition on Matched Formulas}
      \label{sec:matched} 
		
      In this section we shall describe the results of experiments
      we have performed on matched formulas. In particular we were
      interested in phase transition of \(k\)-CNF formulas with
      respect to the property ``being matched'' depending on the ratio
      of the number of clauses to the number of variables. We
      will also compare the results with the theoretical bound proved
      in~\cite{Franco2003177} (see Theorem~\ref{thm:fg-matched}).

      Note that the graphs in \(\mathcal{G}^k_{m,n}\) correspond to
      incidence graphs of \(k\)-CNFs on \(n\) variables and \(m\)
      clauses. In particular, the probability that a random formula
      \(\varphi\in\mathcal{M}^k_{m,n}\) is matched is the same as the
      probability that a random bipartite graph
      \(G\in\mathcal{G}^k_{m,n}\) admits a matching of size \(m\). In
      the experiments we were working with random bipartite graphs and
      we identified them with random formulas. The difference between
      a random formula \(\varphi\) and a random bipartite graph \(G\)
      is in polarities of variables which have no influence on
      whether the formula is matched or not.

      \begin{figure}[htb]
         \centering
         \includegraphics[width=0.99\textwidth]{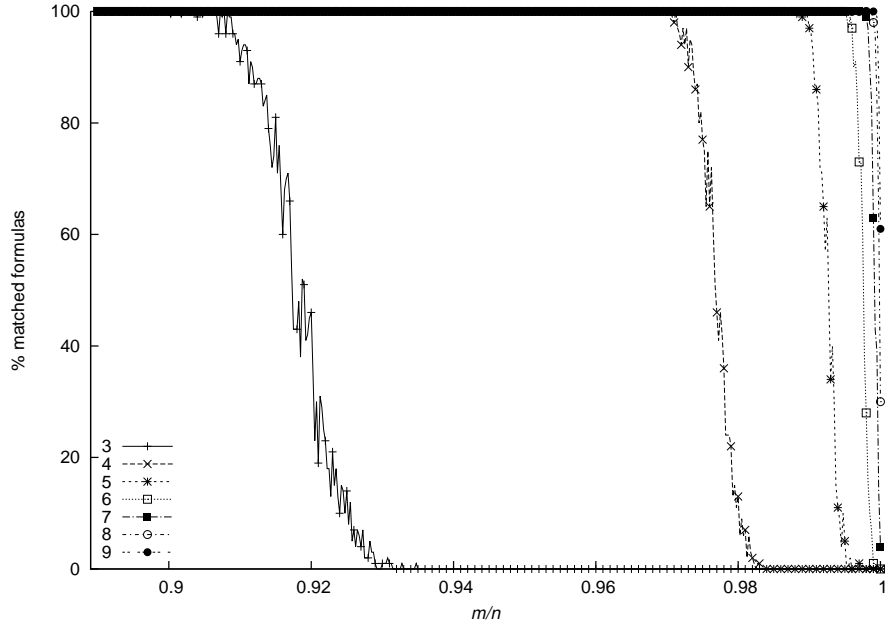}
         \caption{Results of experiments on random graph
            \(\mathcal{G}^k_{m,n}\) with \(n=4000\) and \(k=3,
               \dots, 9\). The horizontal axis represents the ratio \(\frac{m}{n}\). The vertical axis
            represents the percentage of graphs which admit matching of size \(m\). For each
            \(k\) and \(m\) we have generated a \(1000\) random
            graphs from \(\mathcal{G}^k_{m,n}\).}
         \label{fig:matched_more} 
      \end{figure}

      \begin{table}[htb]
         \caption{Phase transition intervals of matched formulas as
                     two values \emph{high} and \emph{low}.
                     We provide only \emph{low} value for \(k\geq 7\),
                     because the \emph{high} value was \(1\) in this
                     case for all configurations. }
         \resizebox{\textwidth}{!}{%
            \begin{tabular}{r||S|S|S|S|S|S|S|S|S|S|S|S}
               & \multicolumn{2}{c |}{3-CNF} & \multicolumn{2}{c |}{4-CNF} & \multicolumn{2}{c |}{5-CNF} & \multicolumn{2}{c |}{6-CNF} & {7-CNF} & {8-CNF} & {9-CNF} & {10-CNF} \\
               \(n\) & {low} & {high} & {low} & {high} & {low} & {high} & {low} & {high} & {low} & {low} & {low} & {low}    \\ \hline\hline
               100  & 0.85   & 0.98    & 0.95    & 1  & 0.97   & 1   & 0.98   & 1   & 1  & 1   & 1   & 1   \\
               200  & 0.88   & 0.96    & 0.96    & 0.99   & 0.98   & 1   & 0.99   & 1   & 1  & 1   & 1   & 1   \\
               500  & 0.89   & 0.95    & 0.96    & 0.99   & 0.99   & 1   & 1  & 1   & 1  & 1   & 1   & 1   \\
               1000 & 0.895 & 0.939  & 0.97    & 0.989 & 0.986 & 0.999  & 0.99   & 0.995  & 0.997 & 0.998  & 0.999  & 0.999  \\
               2000 & 0.903 & 0.9325 & 0.97    & 0.985 & 0.988 & 0.9965 & 0.995 & 0.9995 & 0.998 & 0.9985 & 0.9995 & 0.9995 \\
               4000 & 0.909 & 0.929  & 0.9715 & 0.982 & 0.99   & 0.995  & 0.995 & 0.992  & 0.998 & 0.999  & 0.9995 & 0.9995
            \end{tabular}%
         }
         \label{tab:result_matchin} 
         
      \end{table}

      In our experiments we considered values of number of variables
      \(n=100\), 200, 500, 1000, 2000, 4000 and \(k=3, 4, \dots, 10\).
      For each such pair \(n\), \(k\) we have generated \(1000\)
      random graphs \(G\in\mathcal{G}^k_{m,n}\) for ratio \(\frac{m}{n} =0.64, 0.65, \dots, 1\).
      Figure~\ref{fig:matched_more} shows the graph with the results of
      experiments for value \(n=4000\). The graph contains a different
      line for each value of \(k=3,
         \ldots, 9\) which shows the percentage of graphs which admit matching of size \(m\) among the generated random graphs
      depending on ratio \(\frac{m}{n} =0.64, 0.65, \dots, 1\).
      The complete results of the
      experiments are shown in Table~\ref{tab:result_matchin}. For
      each value of \(k\) we distinguish two values \emph{high} and
      \emph{low} where only \(1\%\) of the graphs generated in \(\mathcal{G}^k_{m,n}\)
      with \(\frac{m}{n}\geq \text{\emph{high}}\) admit matching of
      size \(m\), and on the other hand \(99\%\) of the graphs
      generated in \(\mathcal{G}^k_{m,n}\) with
      \(\frac{m}{n}<\text{\emph{low}}\) admit matching of size \(m\).

      We can see that for higher values of \(n\) the interval
      \([\text{\emph{low}}, \text{\emph{high}}]\) gets narrower
      and we can thus claim that the property ``being matched'' indeed
      exhibits a phase transition phenomenon. Moreover we can say that
      the average of values \emph{low} and \emph{high} limits to
      the threshold of this phase transition. We can see that the
      threshold ratio for \(k=3\) is around \(0.92\) which is much
      higher than the theoretical bound \(0.64\) from~\cite{Franco2003177} (see Theorem~\ref{thm:fg-matched}). In all
      configurations with \(k\geq 7\) the \emph{high} value was \(1\)
      while the \emph{low} value was close to \(1\) as well. Thus in
      the experiments we made with \(k\geq 7\) even in the case
      \(m=n\) almost all of the randomly generated graphs admitted matching of size \(m\).

   \section{Bounded Biclique Cover Heuristic}
      \label{sec:biclique} 

      The class of biclique satisfiable formulas form a natural
      extension to the class of matched formulas. This class was
      introduced by Szeider~\cite{Szeider2005}, where the author showed
      that it is NP-complete to decide whether a given formula
      \(\varphi\) is biclique satisfiable. Recall that this decision
      is equivalent to checking if the incidence graph \(I(\varphi)\)
      has a bounded biclique cover. In this section we shall describe a
      heuristic algorithm for finding a bounded biclique cover. The
      algorithm we introduce is incomplete, which means that it does
      not necessarily find a bounded biclique cover if it exists, on
      the other hand the algorithm runs in polynomial time.
      
      \subsection{Description of Heuristic Algorithm}
    \label{ssec:biclique_description}
      Our heuristic approach is described in
       Algorithm~\ref{alg:heuristika}.
       It is based on a heuristic algorithm for finding a smallest
      biclique cover of a bipartite graph described
      in~\cite{conf/hicss/HeydariMSS07}.
      The algorithm expects three parameters. The first two parameters
      are a bipartite graph \(G\) and an integer \(t\) which restricts
      the size of the first part of bounded bicliques used in the
      cover, in other words only bicliques \(S\) satisfying that
      \(|V_v(S)|\leq t\) are included in the cover which is output by
      the algorithm. The last parameter used in the algorithm is the
      strategy for selecting a seed. 
      
         Let $G$ be a bipartite graph $G=(V_v,V_c,E)$. A \emph{seed}
         in \(G\) is a biclique \(S\) which is
         a nonempty subgraph of \(G\) with \(|V_v(S)|=2\) and
         \(V_c(S)\neq\emptyset\). We say that
         \(S\) is a \emph{maximal seed} if there is no seed
         \(S'\) so that \(V_v(S)=V_v(S')\)
         and \(V_c(S)\subsetneq V_c(S')\).
      
      After initializing an empty cover $\mathcal{C}$, the algorithm starts with  a pruning step ({\tt unitGPropagation}) which is used also in the main cycle.
      In this step a simple
      reduction rule is repeatedly applied to the graph \(G\):
      If a vertex \(C\in V_c\) is present in a single
      edge \(\{v, C\}\), then this edge has to be added into the
      cover \(\mathcal{C}\) as a biclique in order to cover \(C\).
	In this case vertices $v$ and $C$ with all edges incident to $v$ are removed from graph $G$.
	  If a vertex \(C\in V_c\) which is not incident
      to any edge in \(E\) is encountered during this process, the heuristic algorithm
      fails and returns an empty cover.

      The algorithm continues with generating a list \(\mathcal{C}\) of all maximal seeds 
      induced by all pairs \(\{v_i,
            v_j\}\subseteq V_v, i < j\).
      The input graph is modified during the
      algorithm by removing edges and vertices. In the following
      description \(G=(V_v, V_c, E)\) always denotes the current
      version of the graph.


      The main cycle of the algorithm repeats while there are some seeds available and \(G\) does not admit a matching of size \(|V_c|\). This is checked by calling
      function \texttt{testMatched} which also adds the matching to \(\mathcal{C}\) if it is found.
      


      The body of the main cycle starts with selecting a seed
      \(S\) by function {\tt chooseSeed}. This choice is based on a given
      strategy. We
       consider three strategies for selecting a seed:
       Strategy $S_{min}$ chooses a seed with the smallest
       second part. Strategy $S_{max}$ chooses a seed with
       the largest second part. And strategy $S_{rand}$ chooses a
       random seed.
      Seed \(S\) is then expanded by repeatedly 
      calling \texttt{expandSeed}. This function selects
      a vertex \(v\in
         V_v\setminus V_v(S)\) which maximizes the size of the second
      part of the biclique induced in \(G\) with left part being
      \(V_v(S)\cup\{v\}\) (the second part is induced to be all the
      vertices incident to all vertices in \(V_v(S)\cup\{v\}\)). The expansion process
      continues while the size of the first part \(V_v(S)\) satisfies
      the restriction imposed by parameter \(t\) and while
      \(S\) is not a bounded biclique (that is while \(2^{|V_v(S)|}\leq
         |V_c(S)|\)).

      If the expansion process ends due to the restriction on the size
      \(|V_v(S)|\) given by \(t\), \(S\) is not necessarily a bounded
      biclique. In this case we use a function \texttt{restrictSeed}
      which simply removes randomly choosen vertices from \(V_c(S)\) so that \(S\) becomes a bounded biclique.

      Once a bounded biclique \(S\) is found, it is removed
      from the graph and it is added to the cover \(\mathcal{C}\).
      This is realized by a function \texttt{removeBiclique} which simply sets \(V_v\gets
         V_v\setminus V_v(S)\), \(V_c\gets
         V_c\setminus V_c(S)\), and \(E\gets E\cap (V_v\times V_c)\).
      Then we call {\tt unitGPropagation} to prune the graph.
      After that function {\tt removeInvalidSeeds} removes from \(\mathcal{S}\) all seeds \(S'\) with
      \(V_v(S')\cap V_v(S)\neq\emptyset\). For remaining seeds \(S'\in\mathcal{S}\) the
      function sets \(V_c(S')\gets V_c(S')\cap V_c\).
      
	  After the cycle finishes the current cover $\mathcal{C}$ is returned.


       \begin{algorithm}[htb]
       	\DontPrintSemicolon
          \KwData{Bipartite graph \(G(V_v, V_c, E)\),
             $t\in\{2,\ldots,\vert V_v\vert\}$ --- maximal size of $|V_v(S)|$ for a biclique \(S\) which we put into the cover and a seeds selection strategy {\tt st}$\in\{S_{min},S_{rand},S_{max}\}$.}
          \KwResult{biclique cover $\mathcal{C}$ of graph $G$ if a heuristic found one, $\emptyset$ otherwise}
          $\mathcal{C}\leftarrow\emptyset$\;
          \lIf(\tcp*[f]{$\mathcal{O}(nm)$}){{\tt
                   unitGPropagation}$(G,\mathcal{C})$
                fails}{\Return{$\emptyset$}}
          $\mathcal{S}\gets\text{\texttt{generateSeeds}}(G)$\tcp*{$\mathcal{O}(n\ell)$}
          \While(\tcp*[f]{$\mathcal{O}(\ell\sqrt{n})$}){$\vert\mathcal{S}\vert>0$ \textup{\textbf{and not}} {\tt testMatched}$(G,\mathcal{C})$}{
             $S\gets\text{\texttt{chooseSeed}}(\mathcal{S}, \text{\texttt{st}})$\tcp*{$\mathcal{O}(n^2)$}
             \While{$\vert V_v(S)\vert<t \wedge  2^{\vert V_v(S)\vert } \leq\vert V_c(S)\vert$}{
                $S\gets\text{\texttt{expandSeed}}(S)$\tcp*{$\mathcal{O}(\ell+m)$}
             }
             \lIf(\tcp*[f]{$\mathcal{O}(\vert V_c(S) \vert)$}){$2^{\vert V_v(S)\vert } \leq\vert V_c(S)\vert$}{
                $S\gets\text{\texttt{restrictSeed}}(S)$}
             $G\gets\text{\texttt{removeBiclique}}(G,S)$\tcp*{$\mathcal{O}(\ell)$}
             $\mathcal{C}\leftarrow\mathcal{C}\cup\{S\}$\;
             \lIf(\tcp*[f]{$\mathcal{O}(nm)$}){{\tt
                   unitGPropagation}$(G,\mathcal{C})$
                fails}{\Return{$\emptyset$}}
             $\mathcal{S}\gets\text{\texttt{removeInvalidSeeds}}(\mathcal{S})$\tcp*{$\mathcal{O}(n\ell)$}
          }
          \Return{$\mathcal{C}$}
          \vspace{1em}
          \caption{An heuristic for checking if there is a bounded
             biclique cover of a bipartite graph \(G=(V_v, V_c, E)\).
             The complexity of each step is noted in comments where we
             consider \(n=|V_v|\), \(m=|V_c|\), and \(\ell=|E|\).}
          \label{alg:heuristika} 
       \end{algorithm}

      
       Let us estimate the running time of our heuristic algorithm~\ref{alg:heuristika}.
       Let us denote \(n=|V_v|\), \(m=|V_c|\), and \(\ell=|E|\) (also corresponds to the length of a formula).
       Generating all seeds requires time \(O(n\ell)\). The main cycle
       will repeat at most $n$ times, because we cannot have more
       bounded bicliques than the number of vertices in $V_v$. In case
       that the second part is bigger then the first one, graph cannot
       be an incidence graph of a matched formula, so checking if a graph
       admits a matching of size \(|V_c|\)
       has constant time complexity if $m > n$. In case that $m \leq n$
       function \texttt{testMatched} will run in
       $\mathcal{O}(\ell\sqrt{n})$~\cite{HK73,LP86}. All other steps within the main
       cycle (including the pruning step) can be performed in time \(O(n\ell)\) and thus the complexity
       of our heuristic is $\mathcal{O}(n^2\ell)$.

      If a nonempty set of bicliques \(\mathcal{C}\) is returned by
      the algorithm, then it
      is a bounded biclique cover of \(G\).
      It should be
      noted that the opposite implication does not necessarily
      hold, if the seeds are chosen badly then the algorithm
      may fail even if there is some
      bounded biclique cover in \(G\). In the next section we aim to
      evaluate our heuristic algorithm experimentally.

   \subsection{Experimental Evaluation of Heuristic Algorithm}
   \label{sec:heur-exp} 

   In this section we shall describe the experiments performed with our heuristic Algorithm~\ref{alg:heuristika} described in
   Section~\ref{ssec:biclique_description}. 

   Algorithm~\ref{alg:heuristika} works with bipartite graphs. We have tested
   proposed heuristic on bipartite  graphs $G$ from the probabilistic space \(G\in\mathcal{G}^k_{m,n}\) with
   $n=100,200$ and with the degrees of vertices
   in the second part being \(k=3, \ldots, 100\).
   This corresponds to formulas in
   \(k\)-CNF for these values.
   We have considered different sizes of the second part given by
   ratios
   $\frac{m}{n}=1,1.01,\ldots,1.5$. The upper
   bound \(1.5\) was chosen because we were mainly interested in bounded 2-biclique cover. For graphs with
   \(\frac{m}{n}> 1.5\) there is no bounded 2-biclique . For comparison, we have
   also performed experiments with unrestricted sizes of bounded
   bicliques and we have tried the three strategies
   $S_{min},S_{rand}$ and $S_{max}$ for selecting a seed.
   In the experiments we checked whether
   Algorithm~\ref{alg:heuristika} found a bounded biclique cover of a given
   random graph generated according to the above mentioned parameters.

   Due to time complexity of Algorithm~\ref{alg:heuristika}
   we have only generated a hundred random
   graphs in \(\mathcal{G}^k_{m,n}\) for each configuration (given by a
   strategy, bound \(t\) on the size of \(V_v(S)\) of each biclique,
   and ratio \(\frac{m}{n}\)).

      \begin{table}[htb]
         \caption{Results of experiments with our heuristic algorithm on graphs with size of second part$|V_v|=100$.
            Each pair of columns \emph{low} and \emph{high} represents a phase
            transition interval. Each row corresponds to one
            strategy. A more detailed explanation can be found in the
         main text.}
         \begin{center}
            \begin{tabular}{r||S|S|S|S|S|S|S|S|S|S|S|S}
               & \multicolumn{2}{c |}{1} & \multicolumn{2}{c |}{1.1} & \multicolumn{2}{c |}{1.2} & \multicolumn{2}{c |}{1.3} & \multicolumn{2}{c |}{1.4} & \multicolumn{2}{c}{1.5} \\
               & {low} & {high} & {low} & {high} & {low} & {high} & {low} & {high} & {low} & {high} & {low} & {high}    \\ \hline\hline
               $S_{min}^{2,3}$  & 4 & 5 & 5 & 6 & 7 & 8 & 9 & 15 & 13 & 24 & 33 & 47   \\
               $S_{rand}^{2,3}$  & 4 & 5 & 5 & 6 & 7 & 8 & 9 & 15 & 13 & 24 & 33 & 47   \\
               $S_{max}^{2,3}$  & 4 & 5 & 5 & 6 & 7 & 8 & 9 & 15 & 14 & 24 & 41 & 87   \\
               $S_{min}^{\infty}$ & 4 & 5 & 5 & 39 & 7 & 40 & 9 & 40 & 39 & 42 & 36 & 44   \\
               $S_{rand}^{\infty}$ & 4 & 5 & 5 & 39 & 7 & 40 & 9 & 40 & 39 & 42 & 36 & 44   \\
               $S_{max}^{\infty}$ & 4 & 5 & 5 & 6 & 7 & 17 & 15 & 20 & 18 & 22 & 20 & 23   \\
            \end{tabular}%
         \end{center}
         \label{tab:result_biclique} 
      \end{table}

   Table~\ref{tab:result_biclique} summarizes the results of our
   experiments. Each row corresponds to a combination of a strategy for
   selecting a seed and a bound imposed on the size of biclique
   (superscript \(2,3\) for bounded 2-biclique cover, \(\infty\) for general bounded biclique cover).
   Each column corresponds to a ratio
   \(\frac{m}{n}\), we have
   included only ratios \(1\), \(1.1\), \(1.2\), \(1.3\), \(1.4\), and
   \(1.5\) in the table. For each configuration we have two bounds \emph{low} and
   \emph{high} on degree \(k\) of vertices in the second part
   \(V_c\) of graph \(G\). Our heuristic algorithm
   succeeded only
   on \(1\%\) of graphs with degree \(k\leq\text{\emph{low}}\) and on
   the other hand it succeeded on \(99\%\) of graphs with degree
   \(k\geq\text{\emph{high}}\).

   We can see that for a bounded 2-biclique cover $S^{2,3}$, the
   strategies \(S_{min}^{2,3}\) and \(S_{rand}^{2,3}\) are never worse
   than \(S_{max}^{2,3}\) and that they even get better for higher
   ratios. This makes \(S_{rand}\) the best strategy for seed size
   restriction \(S^{2,3}\) --- it is easiest to implement and
   randomness means that repeated calls of our heuristic algorithm
   may eventually lead to finding a biclique
   cover. As we can expect, heuristic performs quite well on lower
   values of ratio \(\frac{m}{n}\) and it gets worse on higher values
   of this ratio.
   For general bounded biclique cover the heuristics
   \(S_{rand}^\infty\) and \(S_{min}^\infty\)
   behave very similarly while \(S_{max}^\infty\) is better in most
   cases.


   \begin{figure}[htb]
      \centering
      \includegraphics[width=0.99\textwidth]{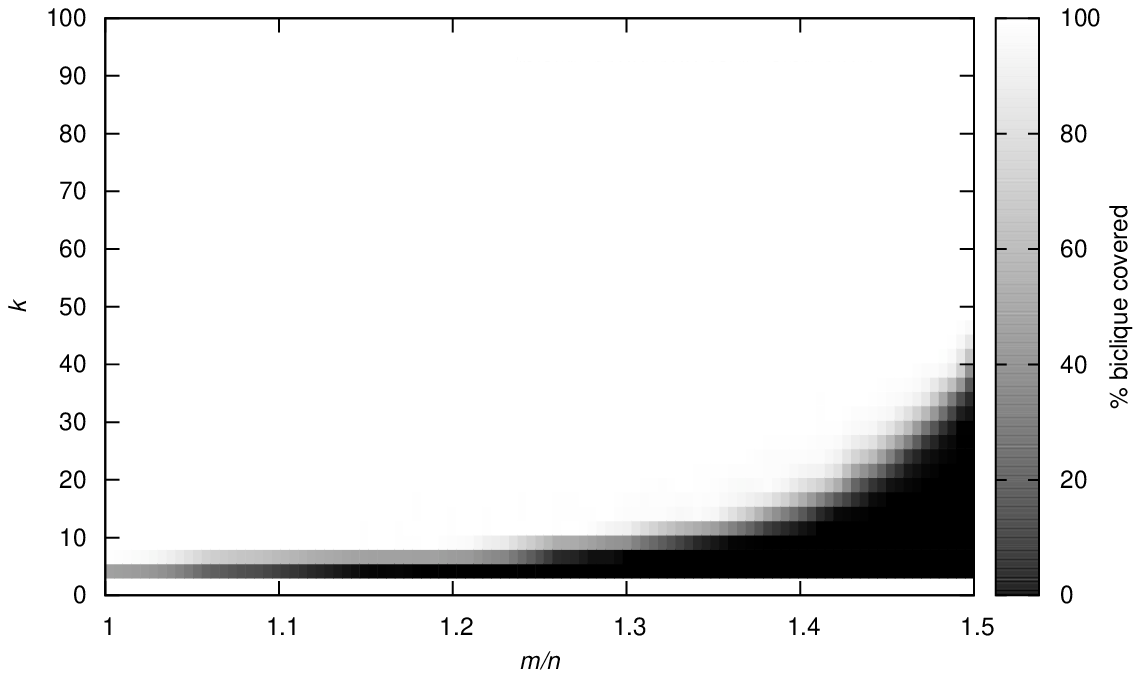}
      \caption{Results of experiments with our heuristic algorithm
         with strategy $S_{rand}^{2,3}$ and $|V_v|=100$. The horizontal axis
         represents the ratio $\frac{m}{n}$.
         The vertical axis represents the degree of vertices $v\in V_c(G)$.
         The more white pixel is, the more random graphs were covered by
         a bounded 2-biclique cover by the algorithm. }
      \label{fig:biclique}
   \end{figure}


   \begin{figure}[htb]
      \centering
      \includegraphics[width=0.99\textwidth]{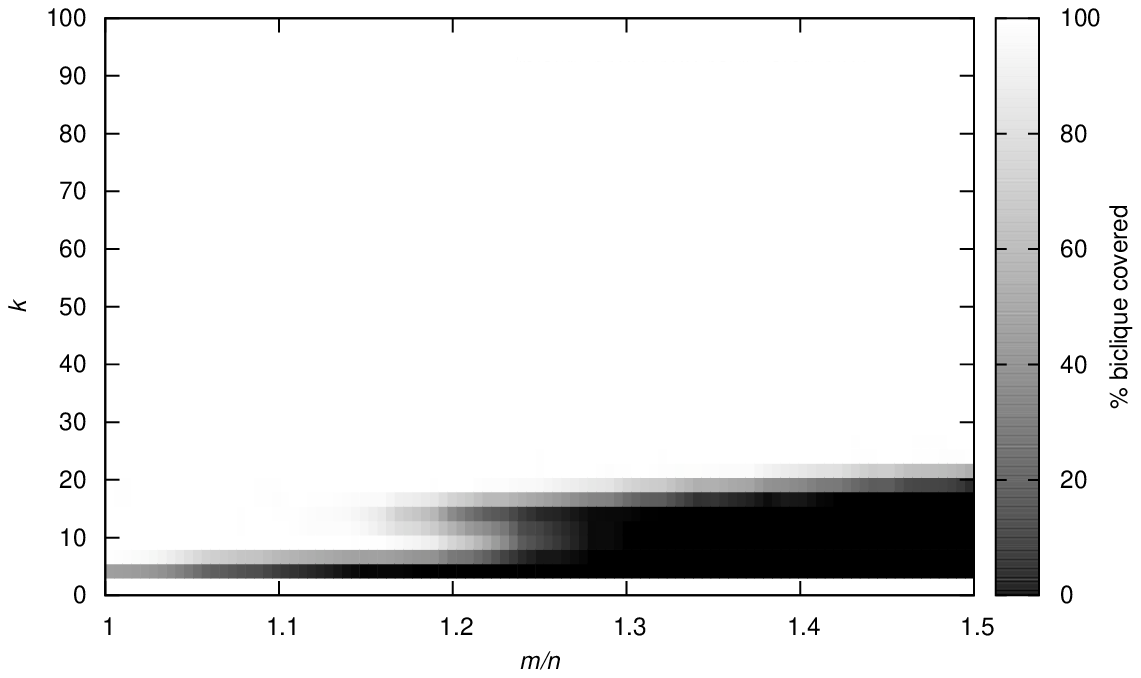}
      \caption{Results of experiments with our heuristic
         algorithm with strategy
         $S_{max}^{\infty}$ and $|V_v|=100$. The horizontal axis
         represents the ratio $\frac{m}{n}$.
         The vertical axis represents the degree of vertices $v\in V_c(G)$.
         The more white pixel is, the more random graphs were covered by
         a general bounded biclique cover by the algorithm.}
      \label{fig:bicliquemax}
   \end{figure}

   We can observe a phase transition behaviour in the results of
   experiments on both strategies \(S^{2,3}\) and
   \(S^\infty\).
   As we can see on Figure~\ref{fig:biclique} and
   Figure~\ref{fig:bicliquemax} there is a phase transition
   \(r_{\frac{m}{n}}\) for a fixed ratio \(\frac{m}{n}\). Most of random graphs
   \(G\in \mathcal{G}_{m,n}^k\) with  \(k \geq r_{\frac{m}{n}}\) have a biclique cover
   and our heuristic algorithm will find it.
   However, since our heuristic is incomplete, it is not clear how
   many random graphs \(G\in\mathcal{G}_{m,n}^k\) with
   \(k \leq r_{\frac{m}{n}}\) have
   biclique cover. 

   In case of strategies with \(S^{2,3}\) the most interesting case is
   when $\frac{m}{n}\leq 1.4$.
   As the ratio \(\frac{m}{n}\) gets close to \(1.5\) we can expect
   smaller percentage of graphs \(G\in \mathcal{G}_{m,n}^k\) having a bounded
   2-biclique cover, hence our heuristic algorithm fails to find one
   in most cases.

   Strategy \(S^\infty\) behaves very similarly to \(S^{2,3}\) but it doesn't have
   an upper limit to phase transition. As we can see, there is an
   interesting phenomenon on Figure~\ref{fig:bicliquemax} between
   \(1.15\) and \(1.25\). The strange shift is caused by using bigger
   bicliques by the algorithm.

   As we can see from Table~\ref{tab:result_biclique}, for ratios
   \(\frac{m}{n}\) smaller than \(1.4\) it is better to use the
   algorithm with a heuristic for finding a bounded \(2\)-biclique cover. For
   bigger ratios \(\frac{m}{n}\) it is better to use a heuristic for general bounded biclique cover. It would be
   also interesting to perform more experiments with bounded 3-biclique covers
   and observe if a similar phenomenon will occur on strategy
   $S^\infty$.

   \begin{table}[htb]
   \caption{Average running time (in $\mu s$) of experiments with our heuristic algorithm. Each column represents }
   \begin{center}
               \resizebox{\textwidth}{!}{%
               \begin{tabular}{r||S|S|S|S|S|S|S|S|S|S|S|S|S|S|S|S|S|S}
                  & \multicolumn{3}{c |}{1} & \multicolumn{3}{c |}{1.1} & \multicolumn{3}{c |}{1.2} & \multicolumn{3}{c |}{1.3} & \multicolumn{3}{c |}{1.4} & \multicolumn{3}{c}{1.5} \\
                                     & 3 & 10 & 20 & 3 & 10 & 20 & 3 & 10 & 20 & 3 & 10 & 20 & 3 & 10 & 20 & 3 & 10 & 20  \\ \hline\hline
                  $S_{rand}^{2,3}$   & 3.01 & 4.66 & 1.18 & 3.01 & 6.03 & 34.25 & 3.01 & 7.83 & 51.26 & 3.01 & 9.09 & 61.68 & 3 & 10.34 & 69.61 & 3 & 11.8 & 77.25\\
                  $S_{max}^{\infty}$ & 3.01 & 4.78 & 1.18 & 3    & 6.47 & 19.35 & 3    & 7.95 & 30.44 & 3.01 & 8.99 & 38.88 & 3 & 9.96  & 44.9  & 3 & 11   & 49.58\\
               \end{tabular}%
               }
            \end{center}
   \label{tab:time}
   \end{table}

   Average runtime of experiments on our heuristic can be seen in
   Table~\ref{tab:time}. For 3-CNF it has the same runtime for both
   strategies and all ratios of $\frac{m}{n}=1,\ldots,1.5$. This
   is because quite often an isolated vertex $v\in V_c$ was created
   during the work of Algorithm~\ref{alg:heuristika}. Which means the
   algorithm failed quickly in many cases. Runtime of strategy
   $S^{2,3}$ which uses bounded bicliques in cover is much worse than
   unbounded strategy $S^\infty$. For $k$-CNF as $k$ grow, the
   difference gets bigger. Its because with unbounded strategy
   $S^\infty$ we admit bigger bicliques in cover and hence our
   heuristic Algorithm~\ref{alg:heuristika} will run fewer iterations
   of the main cycle and succeeds or fails faster than $S^{2,3}$.

  \section{Bounded Biclique SAT Encoding}
  \label{sec:sat-encoding}

  We shall first describe the encoding of the
  problem of checking if a bipartite graph has a bounded biclique cover into SAT,
  then we will describe and evaluate the experiments we have
  performed to compare this approach with Algorithm~\ref{alg:heuristika}. We will also describe the environment we have used to
  run the experiments.

  \subsection{Description of SAT Encoding}
  \label{ssec:encoding}

  A valid biclique $B$ of a bipartite graph $G$ is a complete bipartite
  subgraph $B$ of the bipartite graph $G$ which follows the
  restriction on the size of the second partition. In particular, we
  require $|V_c(B)|<2^{|V_{v}(B)|}$.
  Let us consider a bipartite graph $G$ and \(k\geq 1\), let us define
  \begin{align*}
  \mathcal{B}_k=\{B\mid \text{\(B\) is a valid biclique of
        \(G\) with
        \(|V_{v}(B)|\leq k\)}\}\text{.}
  \end{align*}
  We also denote $\mathcal{B}_\infty$ the set of all bounded bicliques
  within the bipartite graph \(G\) without restriction on the
  size of \(V_{v}\). We would use set of bicliques \(\mathcal{B}_k\) to check existence of a bounded k-biclique cover and \(B_\infty\) to check existence of a general bounded biclique cover. We will encode problem of bounded ($k$-)biclique cover on a bipartite graph \(G=(V_v,V_c,E)\).
  Let us fix \(\mathcal{B}_k\) where \(k\)
  is either a natural number, or \(\infty\) and let us describe
  formula \(\psi\) for a given graph $G$.
  With each biclique $B\in\mathcal{B}_k$ we associate a new variable $x_B$.
  Every assignment of boolean values to variables \(x_B\),
  \(B\in\mathcal{B}_k\) then specifies a set of bicliques. We
  want to encode the fact that the satisfying assignments of \(\psi\) exactly
  correspond to bounded biclique covers of \(G\).
  To this end we use the following constraints:
  \begin{itemize}
     \item For each vertex $v\in V_v$ we add to \(\psi\) an at-most-one
        constraint on variables \(x_B\), \(v\in V_{v}(B)\). This
        encodes the fact that the first partitions of bicliques in the cover have to be
        pairwise disjoint. We use a
        straightforward representation of the at-most-one constraint with
        a quadratic number of negative clauses of size \(2\).
     \item For each clause \(C\in V_c\) we add to \(\psi\) a clause
        representing an at-least-one
        constraint on variables \(x_b\), \(C\in V_c(B)\). This encodes
        the fact that each vertex of second partition belongs to a biclique in the cover.
  \end{itemize}

%

  \subsection{Experimental Evaluation of Heuristic Algorithm}
  \label{ssec:sat-exp-results}

  We can see that the number of variables in our encoding is equal to
  the number of all valid bicliques within the bipartite graph
  \(G\). If we consider
  bicliques in \(\mathcal{B}_k\) for a fixed \(k\), then the number of
  valid bicliques \(|\mathcal{B}_k|\) is polynomial in the size of
  \(\varphi\) but it can be exponential in \(i\) (for
  \(\mathcal{B}_\infty\) the number can be exponential in the size of
  \(\varphi\) as well). For this reason we tested the encoding only
  with bicliques in \(\mathcal{B}_2\), thus checking
  bounded 2-biclique cover. For bigger bicliques the running
  times of experiments increased so much that we would not be able to
  repeat the tests enough times for a reasonable number of variables.
  We used the encoding described in Section~\ref{ssec:encoding} to
  check the success rate of our heuristic algorithm on random bipartite graphs and to
  check the phase transition for an existence of bounded biclique cover. We ran the experiments on \(100\) random bipartite graphs 
  \(G\in\mathcal{G}^k_{m,n}\)  with \(n=40\)  for combinations of
  \(k=1, \ldots, 8\) and the size of the second part \(m=rn\) for \(r=1.00,
     \ldots, 1.25\) with step \(0.05\). For each \(k\) we
  tested random graphs only for the ratios around the
  expected phase transition as observed in the
  Table~\ref{tab:result_biclique}.

  \begin{table}[htb]
  \centering
  \caption{Number of bipartite graphs with bounded biclique cover (found by
     Algorithm~\ref{alg:heuristika}/ SAT finished with true/ SAT finished within time limit).
     If the three numbers are
  missing in a cell, no experiments were run with the corresponding
  configuration. In case of black cells
  we expect values \(100/100/100\) and in case of white empty cells we
  expect \(0/0/100\). See also the description within text.}
  \begin{tabular}{c||r|r|r|r|r|r}
  & 1 & 1.05 & 1.1 & 1.15 & 1.2 & 1.25 \\ \hline \hline
  3 & 18/18/100 & \cellcolor{black!15} 0/0/100 &  &  &  & \\
  4 & \cellcolor{black!30}95/95/100 & \cellcolor{black!30}50/56/100 & \cellcolor{black!15} 7/10/100 & \cellcolor{black!15} 0/1/100 & 0/0/100 &   \\
  5 & 100/100/100 & 100/100/100 & \cellcolor{black!30}90/98/98 & \cellcolor{black!30} 52/87/87 & \cellcolor{black!15}10/34/34 &\cellcolor{black!15} 1/1/1  \\
  6 &\cellcolor{black}  & 100/100/100 & 100/100/100 & 100/100/100 & \cellcolor{black!30} 85/99/99 & \cellcolor{black!30} 42/53/53  \\
  7 & \cellcolor{black} & \cellcolor{black} & \cellcolor{black} & 100/100/100 & 99/100/100 & \cellcolor{black!30} 89/90/90   \\
  8 & \cellcolor{black} & \cellcolor{black} & \cellcolor{black} &  \cellcolor{black}& \cellcolor{black} & 99/100/100
  \end{tabular}

  \label{tab:cmp_res}
  \end{table}

  \begin{table}[htb]
  \centering
    \caption{Average runtime of Algorithm~\ref{alg:heuristika}/average runtime of SAT on
       encoding in seconds.}

  \begin{tabular}{c||r|r|r|r|r|r}
  & 1 & 1.05 & 1.1 & 1.15 & 1.2 & 1.25 \\ \hline \hline
  3 & 0.005/0.014 &  0.005/0.01 &  &  & & \\
  4 & 0.005/1.7 & 0.006/25 & 0.006/21 & 0.005/13 & 0.006/7 & \\
  5 & 0.005/0.2 & 0.005/0.98 & 0.006/11 & 0.006/216 & 0.006/993 & 0.005/4589\\
  6 &  & 0.006/0.4 & 0.006/2.4 & 0.006/50 & 0.006/838 & 0.006/3646\\
  7 &  &  &  & 0.006/73 & 0.006/166 & 0.007/1666\\
  8 &  &  &  &  &  & 0.006/272
  \end{tabular}

  \label{tab:cmp_time}
  \end{table}

  \begin{table}[htb]
    \centering
    \caption{Average/maximum ratio between running time of
       Algorithm~\ref{alg:heuristika} and running
       time of SAT on encoding in seconds. }
  \begin{tabular}{c||r|r|r|r|r|r}
  & 1 & 1.05 & 1.1 & 1.15 & 1.2 & 1.25 \\ \hline \hline 
  3 & 0.52/2.16 & 0.58/2.39 &  &  & & \\
  4 & 0.13/1.05 & 0.02/0.40 & 0.002/0.06 & 0.001/0.005 & 0.001/0.009 & \\
  5 & 0.25/1.13 & 0.06/1.16 & 0.007/0.12 & 0.002/0.096 & $8\cdot 10^{-5}$/0.0008 & $6\cdot 10^{-7}$/$6\cdot 10^{-7}$\\
  6 &           & 0.12/0.88 & 0.023/0.30 & 0.003/0.095 & 0.0002/0.007 & $8\cdot 10^{-5}$/0.0007\\
  7 &           &           &            & 0.006/0.086 & 0.001/0.039 & 0.00018/0.0015\\
  8 &           &           &            &             &               & 0.00026/0.0028
  \end{tabular}

  \label{tab:cmp_ratio_time}
  \end{table}

  The results of experiments are contained in tables~\ref{tab:cmp_res}
  to~\ref{tab:cmp_ratio_time}. All these tables have a similar
  structure. Each cell represents a single configuration (row
  corresponding by
  a value of \(k\) and column corresponding to the ratio \(m/n\) where \(m\) denotes the
  number of clauses and \(n=40\) denotes the number of variables). In
  Table~\ref{tab:cmp_res} each cell contains three numbers separated
  with slashes. The first is the number of instances (out of \(100\))
  on which Algorithm~\ref{alg:heuristika} successfully found a bounded biclique
  cover. The second is the number of instances on which the SAT solver
  successfully solved the encoding and answered positively. The third
  is the number of instances on which the SAT solver finished within
  time limit which was set to 4 hours for each instance. In some cells
  the values are missing, for these configurations we did not run any
  experiments, because they are far from the observed phase transition
  (see Table~\ref{tab:result_biclique}). In case of
  black colored cells we assume that the results would be
  \(100/100/100\), in case of white colored cells we assume that the
  results would be \(0/0/100\). The gray colored cells mark the
  borders of observed phase transition intervals of existence 
  of bounded biclique cover, light gray corresponds to the results
  given by the SAT solver, dark gray to the results given by the
  heuristic which form an upper bound on the correct values.
  We can see that in most cases the number of positive
  answers given by Algorithm~\ref{alg:heuristika} is close to the
  number of positive answers given by the SAT solver. However, there
  are some cases where one of the approaches was more successful ---
  namely in cases of \(k=5, m/n=1.15, k=5\) and \(k=5, m/n=1.2\).
  In the first case  the SAT solver answered on
  \(87\) instances positively while Algorithm~\ref{alg:heuristika}
  answered positively only on \(52\) instances and in the second one
  SAT solver answered on \(34\) positively and
  Algorithm~\ref{alg:heuristika} answered positively only on \(10\)
  instances. However, in the second case we can see that SAT solver run
  over time limit (4 hours) in \(\frac{2}{3}\) cases.

  We also compared runtime of our heuristic algorithm and the SAT
  solver. As we can see in the Table~\ref{tab:cmp_time} our heuristic
  algorithm is much faster in average case. Standard deviation
  of runtime of our heuristic algorithm is around $10^{-2}$ and the
  standard deviation of runtime of SAT solver is up to $10^4$ (where
  we have evaluated the average value and the standard deviation only
  on instances in which the SAT solver finished within the time limit
  4 hours). These values are quite high compared to the running times.
  One of the reasons is perhaps the fact that the
  experiments were not run on a single computer, but on several
  comparable computers (see Section~\ref{ssec:exp-environ} for more
  details). Although in all cases on a single instance, the SAT solver
  and Algorithm~\ref{alg:heuristika} were run on a single computer and
  it makes thus sense to look at the ratio between the running times
  of these two. These are contained in Table~\ref{tab:cmp_ratio_time}.
  We can see that our heuristic is in most cases faster than the SAT
  solver using the encoding described in Section~\ref{ssec:encoding}.
  Only for $k=3$ we can see that the maximum ratio is bigger than
  one. It means that in some cases SAT solver was
  faster than Algorithm~\ref{alg:heuristika}, although not on average.

  We can see from the results that on random \(k\)-CNF formulas Algorithm~\ref{alg:heuristika}
  has a success rate close to the one of the SAT based approach. A big
  advantage of Algorithm~\ref{alg:heuristika} is that it is much
  faster.

  \subsection{Experiments environment}
  \label{ssec:exp-environ}

  Let us say more on the environment in which the experiments were
  run. We used Glucose parallel SAT solver~\cite{glucose,minisat}.
  Our
  experiments were executed on grid computing service
  MetaCentrum NGI~\cite{metavevo}. All experiments were run on a
  single processor machine
  (Intel Xenon, AMD Opteron) with 4 cores and frequency
  2.20GHz-3.30GHz. On each random bipartite graph $G\in\mathcal{G}^k_{m,n}$, Algorithm~\ref{alg:heuristika} and
  the SAT solver were always run on the same computer.
  However, for the same configuration and different formulas, the
  experiments may have run on different computers.
  As we have noted in Section~\ref{ssec:sat-exp-results},
  this could be a reason of significantly high values of standard
  deviation of runtimes.
  The fact that the computer speed varied while the time limit for the
  SAT solver was still the same (4 hours) could have led to situations
  where the SAT solver would not finished, because it was run on a
  slower computer, and could potentially finish had it been run on a
  faster computer.
  We can see in Table~\ref{tab:finished_cases} that most of the
  total 2000 instances finished within an hour, then only 26 finished
  between an hour and 2 hours, only 15 finished between 2 hours and
  3 hours and only 9 finished between 3 hours and 4 hours.
  We can thus expect that the number of the border cases is
  similarly small. We can conclude that the variance in computer speeds had only minor
  influence on the number of SAT calls which finished within the time
  limit.
  \begin{table}[h]
  \centering
  \caption{Number of test cases finished in given interval.}
  \begin{tabular}{c | c | c | c | c | c}
  $<$1h & 1h-2h & 2h-3h & 3h-4h & $>$4h & total\\ \hline \hline
  1912 & 26 & 15 & 9 & 238 & 2200
  \end{tabular}
  \label{tab:finished_cases}
  \end{table}

   \section{Conclusion}
   \label{sec:concl}

   The first result of our paper is that the experimental threshold of
   phase transition of property ``being matched'' of 3-CNFs is around
   0.92 which is much higher than the theoretical lower
   bound 0.64 proved for 3-CNF by J. Franco and A.V.
   Gelder~\cite{Franco2003177}. This can be seen in
   Figure~\ref{fig:matched_more}. Moreover our experiments suggest
   that for $k\geq 6$ almost all formulas in \(k\)-CNF are matched (if
   they have at most as many clauses as variables).

   We have also proposed a heuristic algorithm
   for finding a bounded ($k$-)biclique cover of a incidence graph
   $I(\varphi)$ of a given formula \(\varphi\). In other words the
   algorithm tries to
   decide if $\varphi$ is biclique satisfiable. We suggested three
   different strategies for selecting a seed in our heuristic and
   compared them. We can deduce from figures~\ref{fig:biclique}
   and~\ref{fig:bicliquemax} that the success rate of our
   heuristic algorithm exhibits a phase transition phenomenon similar
   to the case of matched formulas. The exact values are shown in
   Table~\ref{tab:result_biclique}.
   Our results suggest that it is better to use
   Algorithm~\ref{alg:heuristika} to find a 2-biclique cover using
   strategy \(S_{rand}\)
   for ratios \(\frac{m}{n}\leq 1.4\) where \(m\)
   denotes the number of clauses and \(n\) denotes the number of
   variables in a given formula.
   For higher ratios it is better
   not to restrict the size of the first part of the bicliques in the
   cover and to use strategy \(S_{max}\).

   \begin{table}[htb]
      \centering
      \caption{In this table we compare the phase transitions of
         property ``being matched'' and the success rate of
         Algorithm~\ref{alg:heuristika}. Columns \emph{low} and
         \emph{high} have the same meaning as in tables~\ref{tab:result_matchin} and~\ref{tab:result_biclique}
         }
	  \begin{tabular}{r || S | S | S  | S }
      & \multicolumn{2}{c|}{matched} & \multicolumn{2}{c}{heuristic} \\
      & low & high & low & high \\ \hline \hline
      3-CNF  & 0.909 & 0.929 & 0.9  & 0.93 \\
      5-CNF  & 0.99  & 0.995 & 1.02 & 1.08 
      \end{tabular}
      \label{tab:compar}
   \end{table}

   Table~\ref{tab:compar} presents a comparison of the results experiments on matched formulas with the results of experiments with Algorithm~\ref{alg:heuristika}.
   As we can see, the success rate of
   Algorithm~\ref{alg:heuristika} exhibits a very similar phase
   transition to matched formulas.
   We can see that the \emph{low} bound of the phase transition
   interval in case of matched formulas in \(5\)-CNF is \(0.985\).
   In case of our algorithm the \emph{low} bound of the phase
   transition interval is \(1.02\).
   A formula can be matched only if the ratio \(\frac{m}{n}\) of the
   number of clauses \(m\) to a number of variables \(n\) is at most
   \(1\). According to the results of our experiments a random
   \(k\)-CNF with \(k>5\) is matched with high probability even in
   case the ratio \(\frac{m}{n}\) is \(1\).
   However, for \(7\)-CNFs the \emph{low} value of phase transition
   of our algorithm equals \(1.14\)
   and for $k\geq 10$ it is even more than \(1.3\), which means that
   if \(\varphi\) is a formula in \(10\)-CNF with \(n\) variables and
   at most \(1.3n\) clauses,
   Algorithm~\ref{alg:heuristika} will most likely find a bounded biclique
   cover of the incidence graph of \(\varphi\). These results are
   summarized in Table~\ref{tab:result_biclique}.

   Our heuristic algorithm is not complete, in particular, it
   can happen that a formula is biclique satisfiable, but
   Algorithm~\ref{alg:heuristika} is unable to detect it. It means
   that we can only trust a positive answer of the algorithm.
   We have compared our heuristic with a SAT based approach which can also
   check that a formula is not biclique satisfiable. We can see in
   Table~\ref{tab:cmp_res} that formulas on which
   Algorithm~\ref{alg:heuristika} fails to answer correctly, are
   concentrated around the observed phase transition, and that the
   algorithm answers correctly in most cases for other configurations.
   We can say that the success rate of
   Algorithm~\ref{alg:heuristika} is not far from the complete SAT
   based method. Moreover, as we can see in tables~\ref{tab:cmp_time}
   and~\ref{tab:cmp_ratio_time}, our heuristic is significantly faster than
   a SAT solver on the encoding we have described.

   \section{Future work}
   \label{sec:future} 

   There is still some space to improve our results. 
   
   We can try to develop better heuristics for selecting a seed and other
   steps of our algorithm. For example we can use other
   sizes of bicliques than \(K_{1,1}, K_{2,3}\) and unbounded ones. It
   would also be interesting to test our heuristic on $k$-biclique
   cover for $k>2$. Additionally a deterministic selection heuristic of
   vertices in function {\tt restrictSeed} could improve the success
   rate of our heuristic Algorithm~\ref{alg:heuristika}.
   
   It would be also interesting to find a better SAT encoding of the problem which would allow us to run experiments on bigger instances of input formulas.
   
   The last question is, can be our heuristic algorithm and SAT encoding generalized to var-satisfiability?


\bibliographystyle{splncs04}
\bibliography{mybib}

\end{document}